\documentclass{moriond}

\def\Journal#1#2#3#4{{#1} {\bf #2}, #3 (#4)}

\def\PRL{\em Phys. Rev. Lett.}

\usepackage{caption}
\usepackage{subcaption}

\def\be{\begin{equation}}
\def\ee{\end{equation}}
\def\bea{\begin{eqnarray}}
\def\eea{\end{eqnarray}}

\newcommand{\Photo}{}

\begin{document}
\vspace*{4cm}
\title{OBSERVATION OF SINGLE TOP-QUARK + PHOTON PRODUCTION WITH THE ATLAS DETECTOR}

\author{ M. ALHROOB \footnote{On behalf of the ATLAS Collaboration. \\
\hbox{\scriptsize Copyright 2022 CERN for the benefit of the ATLAS Collaboration. CC-BY-4.0 license.}\
  }}

\address{University of Oklahoma,\\
Norman, Oklahoma, USA}

\maketitle\abstracts{
These proceedings present the first observation of single top-quark production associated with a photon using the full Run2 proton-proton dataset collected by the ATLAS experiment~\cite{Aad:2008zzm} at the LHC.}

\section{Introduction}
Single top-quark production with an associated vector boson is a rare process predicted by the Standard Model (SM) and fundamental for probing the top-quark electroweak couplings. These processes play a role in constraining non-resonant contributions of Physics Beyond the SM parametrised in the effective field theory framework. The ATLAS and CMS collaborations have observed single top-quark production processes with both a W boson (tW)~\cite{ATLAS:2015igu,CMS:2014fut} and a Z boson (tZq)~\cite{CMS:2018sgc,ATLAS:2020bhu}. The latest CMS measured tZq cross-section is 87.9 fb with 12\% uncertainty~\cite{CMS:2021ugv}. The CMS collaboration reported evidence for the associated single top-quark production with a photon ($tq\gamma$) with a 4.3 $\sigma$ significance using a partial Run2 dataset of 35.9 fb$^{-1}$ of proton-proton collisions collected at 13 TeV with a measured $tq\gamma$ cross-section of 115 fb $\pm$ 30\%~\cite{CMS:2021ugv}. 

In these proceedings, we report the observation of the $tq\gamma$ production process made by the ATLAS collaboration using the full Run2 13 TeV proton-proton dataset of 139 fb$^{-1}$ collected by the ATLAS detector at the LHC. 

\section{Signal definition and fiducial cross-sections}
\begin{figure}[htb]
  \centering
    \includegraphics[width=.4\textwidth]{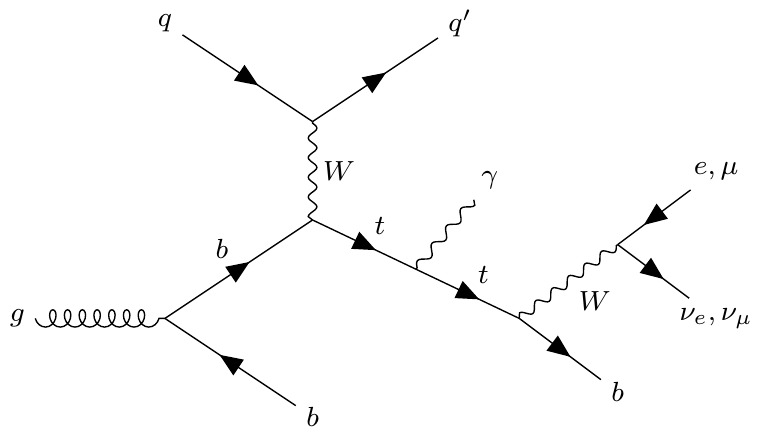}
    \caption{Feynman diagram at leading order in QCD for $tq\gamma$ production with semi-leptonic top-quark decay~\protect\cite{ATLAS-CONF-2022-013}.}   
     \label{Feyn_tqy}
\end{figure}

In the search for the $tq\gamma$ production process~\cite{ATLAS-CONF-2022-013}, the ATLAS collaboration considered only semileptonic top-quark decays. The photon can be radiated from the either the top quark or other hard scattered particle in the interaction, such as an initial state quark. The photon can be also radiated by the charged particles from the top-quark decay. Figure~\ref{Feyn_tqy} represents an example of one of the Feynman diagrams where a photon is radiated from the top quark, with subsequent semileptonic top-quark decay. The signal signature consists of a photon, an electron or muon, missing transverse momentum from the neutrino, a $b$-tagged jet, and a forward jet due to t-channel production.

The $tq\gamma$ production cross-section is measured in fiducial phase space at the parton level, excluding photons radiated from the top-quark decay products. A particle-level fiducial cross-section is also measured, including photons radiated from the top-quark decay products, covering the signal phase space. The parton-level fiducial cross-section is defined by requiring at least an isolated photon with $p_{T}\ge$ 20 GeV, $|\eta|\le$2.5 and $\Delta R >$ 0.4 from final state hard-scattering charged partons. Also, a charged lepton is required with $|\eta|\le$2.5. The parton-level fiducial cross-section is multiplied by the branching fraction of the top-quark decaying semileptonically, $\sigma_{tq\gamma}\times{\cal{B}}(t\rightarrow\ell\nu b)$, is $406^{+25}_{-32}$ fb where the uncertainty is due to varying the PDF and renormalisation and factorisation scales.

Similarly, the fiducial cross-section is calculated at the particle level, including photons radiated from the top-quark decay products. The photon production from final-state particles is estimated using the traditional SM $t$-channel process with photon production simulated through the parton shower. The fiducial volume is defined by requiring at least one isolated photon with $p_{T}\ge$ 20 GeV and $| \eta| <$ 2.37, one electron or muon with $p_{T}\ge$ 25 GeV and $| \eta| <$ 2.5, at least one b-jet with $p_{T}\ge$ 25 GeV and $| \eta| <$ 2.5, at least one neutrino, excluding those coming from the hadronic decay. Jets within $\Delta R =$ 0.4 from a photon or a lepton are removed. Events are removed if a photon is within $\Delta R =$ 0.4 from a lepton or a jet. The calculated fiducial particle-level cross-section multiplied by the branching fraction of the top-quark decaying semileptonically, $\sigma_{tq\gamma}\times{\cal{B}}(t\rightarrow\ell\nu b)+\sigma_{(t\rightarrow\ell\nu b\gamma)q},$ is $207^{+26}_{-11}$ fb with $20\%$ contribution estimated from events with photons radiated from top-quark decay products. The uncertainty is due to varying the PDF and renormalisation and factorisation scales.

\section{Event selection}
Events are selected using a single lepton ($e$ or $\mu$) trigger and are required to have exactly one reconstructed well-isolated lepton with $p_{T}>27$ GeV and $|\eta|<$ 2.47 for electrons and $|\eta|<$ 2.5 for muons. At least one well-isolated photon with $p_{T} >$ 20 GeV and $|\eta|<$ 2.37 is required. Electrons and photons are rejected if they are within 1.37 $< |\eta|<$ 1.52. Due to the presence of a neutrino, a minimum missing transverse momentum, $E^{\rm miss}_{\rm T}$, is required to be greater than 30 GeV. To reduce the contribution of the $Z\gamma$ background due to electrons faking photons, $|m_{e\gamma} - 90| > $10 GeV criteria is required. Also, events are required to have at least one jet with $p_{T}>$25 GeV within $|\eta|<4.5$, in which exactly one jet is b-tagged with 70\% efficiency. Events with additional loose b-jets (tagged with 85\% efficiency) are rejected. Two signal regions (SRs) are defined based on the number of forward jets (fj) ($4.5>|\eta|>$2.5): with $\ge 1$ fj SR, which has the most number of signal events, and with $= 0$ fj SR. The dominant background processes are $t\bar{t}\gamma$ and $W\gamma$ which are measured using two dedicated control regions (CRs). Table~\ref{tab:SR_CR} shows the SRs and CRs designed to extract the signal and control the dominant background processes.

\begin{table}[htb]
\centering
\begin{tabular}{|c| c| c |c |c|} 
 \hline
     & $\ge$ 1 fj SR & 0 fj SR & $t\bar{t}\gamma$ CR & $W\gamma$ CR\\
 \hline
     Leptons ($p_{T}\ge$ 27 GeV) & = 1 & = 1 & = 1 & = 1 \\ 
\hline
 $E^{{\rm miss}}_{T}$ & $\ge$ 30 GeV & $\ge$ 30 GeV & $\ge$ 30 GeV  &  $\ge$ 30 GeV \\
 \hline
 Photons ($p_{T}\ge$ 20 GeV) & $\ge$ 1 & $\ge$ 1 & $\ge$ 1 & $\ge$ 1 \\ 
 \hline
 $|m_{e\gamma}-91.2|$ & $\ge 10$ GeV & $\ge 10$ GeV & $\ge 10$ GeV  & $\ge 10$ GeV  \\ 
 \hline
 forward jets ($2.5 < |\eta| < 4.5$ ) & $\ge$ 1 & {\color{red}$=0$} & $\ge 0$   & $\ge 0$ \\
 \hline
 b-jets (70\% WP)  & = 1 & = 1 & =1  &  {\color{red}= 0} \\
 \hline
Additional b-jets (85\% WP)  & = 0 & = 0 & {\color{red}$\ge$ 1}  &  {\color{red}$\ge$ 1} \\
 \hline
\end{tabular}
 \caption[]{A summary of the signal and control regions. In red, the selection that leads to the orthogonality with other regions.}
  \label{tab:SR_CR}
\end{table}

The contribution of processes with electrons faking photons, mainly dilepton $t\bar{t}$ and Z+jets, is estimated by comparing data and MC using the tag-and-probe method $N(Z\rightarrow e(e\rightarrow\gamma))/N(Z\rightarrow e^+e^-)$. An enriched region with $e\rightarrow\gamma$ fakes is selected by inverting $m_{e\gamma}$ or $m_{e^+e^-}$ ( e.g $|m_{e\gamma}-90| < 20$ GeV), $E^{\rm miss}_{\rm T}$ cuts and by requiring no $b$-tagged jet. This estimate is validated in a region with similar requirements but with $\ge 1$ $b$-tagged jet requirement. The contribution of processes with hadrons faking photons, mainly lepton+jets $t\bar{t}$, is estimated using the ABCD method. Where the ABCD regions' definitions are based on photon identification and isolation criteria. The correction factors are $p^{\gamma}_{T}$ and $\eta^{\gamma}$ dependant.

\section{Signal and background separation}
\begin{figure}[htb]
    \centering
  \begin{subfigure}[a]{.24\textwidth}
    \includegraphics[width=\textwidth]{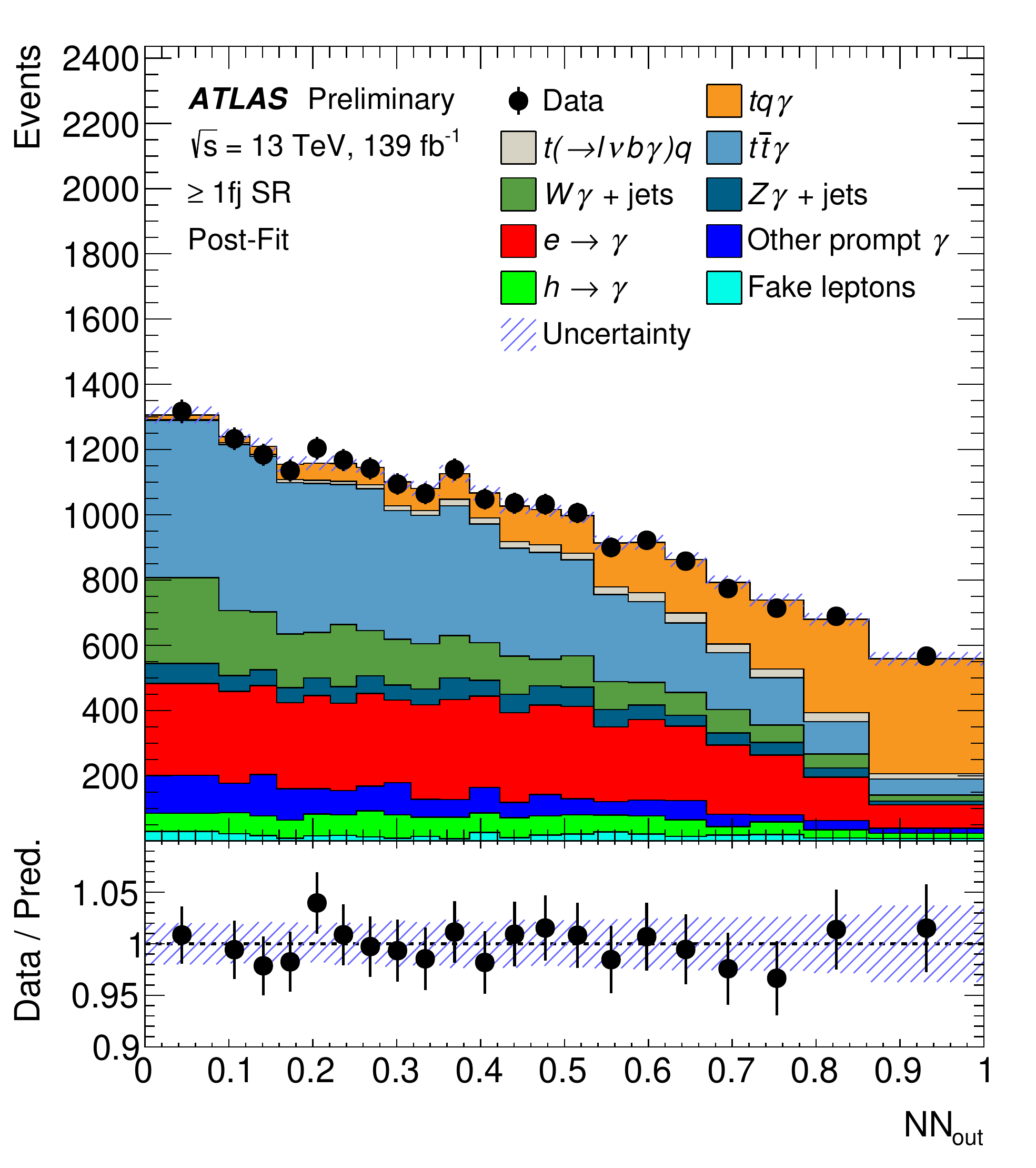} 
    \caption{}
    \label{fig:NN_SR1}
  \end{subfigure}
    \begin{subfigure}[a]{.24\textwidth}
    \includegraphics[width=\textwidth]{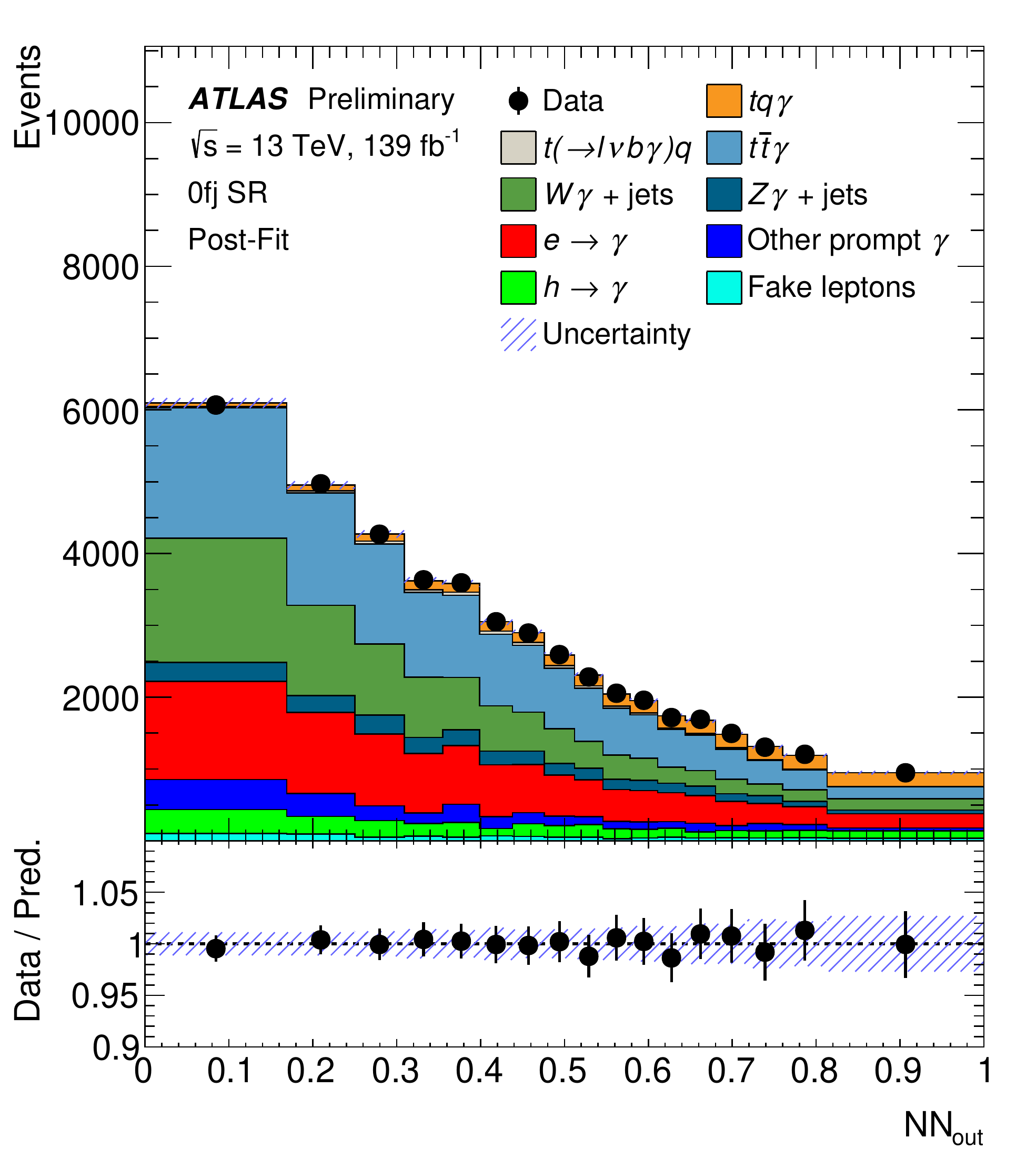} 
    \caption{}
    \label{fig:NN_SR2}
  \end{subfigure}
    \begin{subfigure}[a]{.24\textwidth}
    \includegraphics[width=\textwidth]{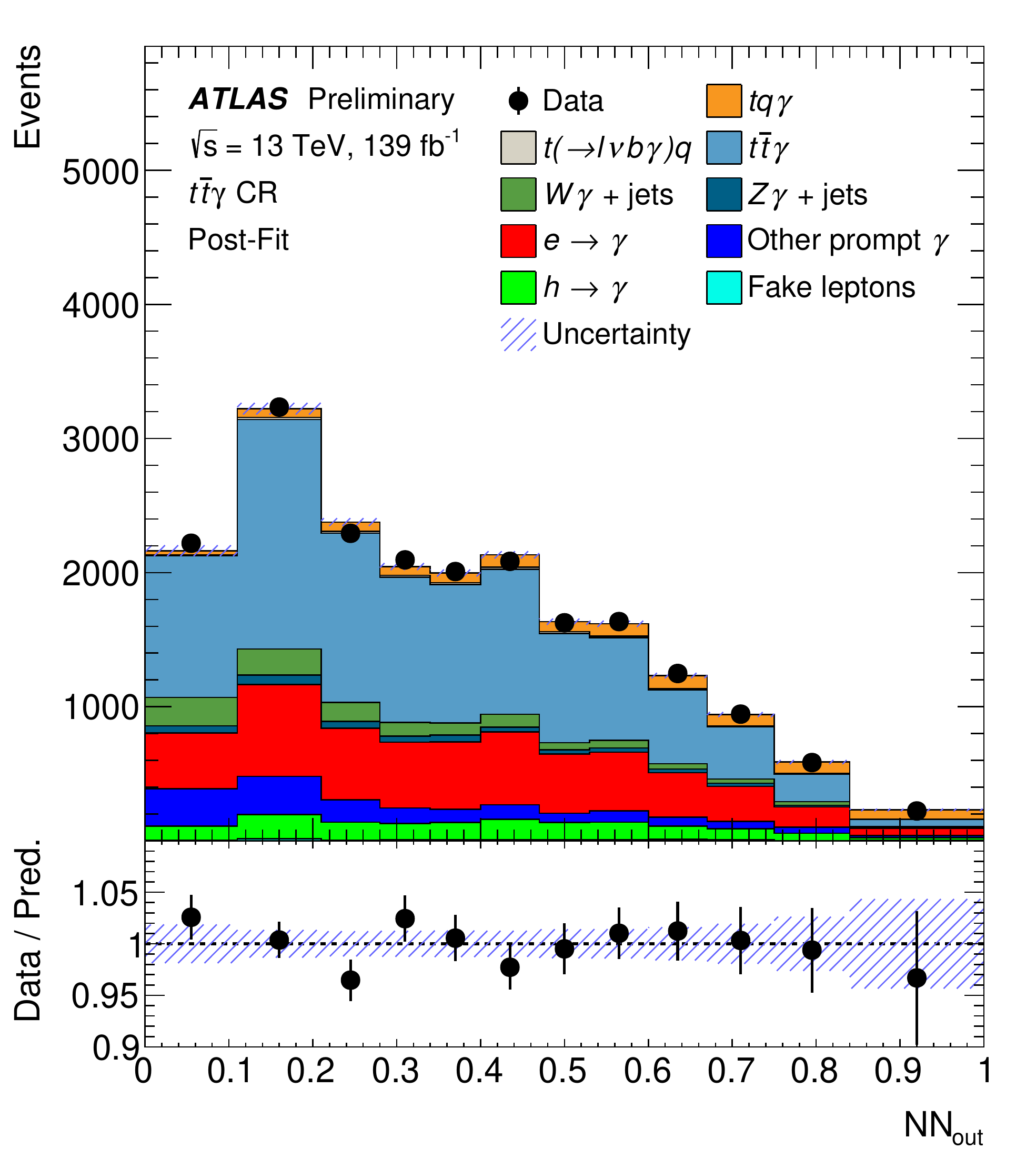} 
    \caption{}
    \label{fig:NN_CR1}
  \end{subfigure}
 \begin{subfigure}[a]{.24\textwidth}
    \includegraphics[width=\textwidth]{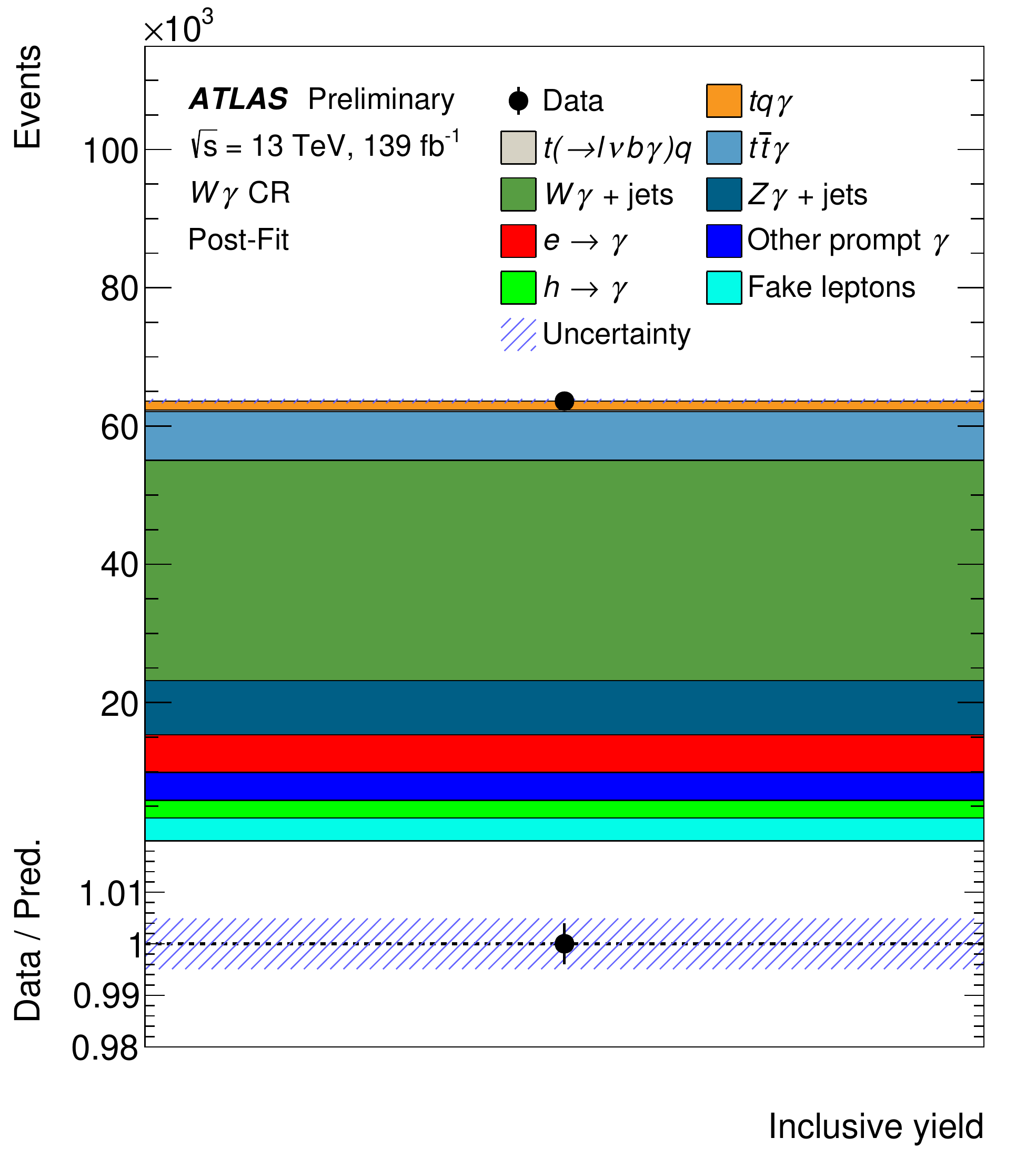} 
    \caption{}
    \label{fig:CR2}
  \end{subfigure}
    \caption{Post-fit distributions of the NN output in the $\ge1$ fj SR (a), 0 fj SR (b), $t\bar{t}\gamma$ CR (d) and the total event yield in $W\gamma$ CR~\protect\cite{ATLAS-CONF-2022-013}.}
     \label{fig:fitted_distributions}
\end{figure}

A deep, fully connected feedforward neural network (NN) is trained for each SR to enhance the signal and background separation. Each NN consists of three hidden layers with 128/8/32 and 128/16/16 hidden nodes for the $\ge 1$ fj and 0 fj SRs, respectively. A Rectified Linear Unit (ReLU) activation function is used for the hidden nodes, while for the output node, a sigmoid function is used. 
For the $\ge 1$ fj SR, 15 well-modelled input variables are selected, while 12 well-modelled input variables are employed for the 0 fj SR, respectively. 
Figure~\ref{fig:NN_SR1} presents the post-fit NN distribution of the $\ge 1$ fj SR, where a clear separation between the signal and background processes is shown.
Figure~\ref{fig:NN_SR2} presents the post-fit NN distribution of the 0 fj SR. The two NNs are also used to construct the NN distribution for the $t\bar{t}\gamma$ CR as shown in~\ref{fig:NN_CR1}, where the NN choice is based on the number of forward jets in the event.

\section{Results}
The $tq\gamma$ cross-section is extracted using a binned maximum-likelihood fit that is performed on the four signal and control regions presented in Figure~\ref{fig:fitted_distributions}. Three NN distributions are used in the two SRs and $t\bar{t}\gamma$ CR, while in the $W\gamma$ CR, a single bin is used. The fit containes three free-floating parameters: one for the signal, one for the $t\bar{t}\gamma$ background and the last for $W\gamma$ background. Other backgrounds and systematic uncertainties are constrained using Gaussian priors.
The significance of the observed signal is 9.1 $\sigma$ compared to the predicted significance of 6.7 $\sigma$.
The measured parton-level fiducial cross-section multiplied by the branching fraction of the semileptonic decay of the top quark is $\sigma_{tq\gamma}\times{\cal{B}}(t\rightarrow\ell\nu b) = 580 \pm 19\, {(\rm stat.)} \pm 63\,{(\rm syst.)}$ fb.
While the measured particle-level fiducial cross-section multiplied by the branching fraction of the semileptonic decay of the top quark is $\sigma_{tq\gamma}\times{\cal{B}}(t\rightarrow\ell\nu b)+\sigma_{(t\rightarrow\ell\nu b\gamma)q} = 287 \pm 8\, {(\rm stat.)} \pm 31\,{(\rm syst.)}$ fb.

The dominant systematic uncertainties affecting the parton-level cross-section are $t\bar{t}\gamma$ modelling with 5.6\%, jets and $E^{\rm miss}_{\rm T}$ with 4\%, background MC statistics with 3.5\%, $tq\gamma$ MC statistics with 3.4\% and $t\bar{t}$ modelling with 3.4\%. Similarly, the dominant systematic uncertainties for the particle-level cross-section are $t\bar{t}\gamma$ modelling with 5.7\%, jets and $E^{\rm miss}_{\rm T}$ with 3.9\%, background MC statistics with 3.5\%, $tq\gamma$ MC statistics with 3.1\% and $t\bar{t}$ modelling with 3.1\%.  

\section{Conclusion}
The ATLAS collaboration at the LHC has observed the $tq\gamma$  process with a significance of 9.1 $\sigma$ using the full Run2 dataset of 139 fb$^{-1}$ collected at 13 TeV, compared to the predicted significance of 6.7 $\sigma$~\cite{ATLAS-CONF-2022-013}.
The measured fiducial cross-section at parton-level multiplied by the branching fraction of the semileptonic decay of the top quark is 580 fb $\pm$ 11\%. In comparison, the measured particle-level fiducial cross-section multiplied by the branching fraction of the semileptonic decay of the top quark is 287 fb $\pm$ 11\%. These measured cross-sections are within 2.5 $\sigma$ and 1.9 $\sigma$ from the SM predictions for the parton-level and particle-level fiducial cross-sections, respectively.

\end{document}